\documentclass[a4paper,UKenglish,cleveref, autoref, thm-restate]{lipics-v2021}

\title{Friend or Foe? \\ Anomalous Peers in Monero's P2P Network} 

\titlerunning{Friend or Foe?} 

\author{Yannik Kopyciok}{Technische Universität Berlin, Germany}{kopyciok@tu-berlin.de}{https://orcid.org/0009-0006-4257-4151}{} 
\author{Stefan Schmid}{Technische Universität Berlin, Germany}{schmiste@gmail.com}{https://orcid.org/0000-0002-7798-1711}{}
\author{Friedhelm Victor}{TRM Labs, San Francisco, USA}{friedhelm@trmlabs.com}{https://orcid.org/0000-0001-8329-3133}{}

\authorrunning{Y. Kopyciok, S. Schmid, and F. Victor} 

\Copyright{Yannik Kopyciok, Stefan Schmid, and Friedhelm Victor} 



\relatedversion{}
\relatedversiondetails{Full Version}{https://doi.org/10.48550/arXiv.2509.10214}

\supplement{}
\supplementdetails[subcategory={}, cite={}, swhid={}]{Software}{https://github.com/ykpyck/monero-traffic-analysis}

\acknowledgements{Supported by German Research Foundation (DFG), SPP 2378/ReNO-2 (grant 511099228), 2025-2029.}

\nolinenumbers 

\usepackage{tikz}
\usetikzlibrary{shapes.callouts}
\usetikzlibrary{arrows.meta, positioning, calc}

\EventEditors{Aggelos Kiayias and Maria Kyropoulou}
\EventNoEds{2}
\EventLongTitle{8th Conference on Advances in Financial Technologies (AFT 2026)}
\EventShortTitle{AFT 2026}
\EventAcronym{AFT}
\EventYear{2026}
\EventDate{October 6--9, 2026}
\EventLocation{London, UK}
\EventLogo{}
\SeriesVolume{395}
\ArticleNo{32}

\begin{document}

\maketitle

\newcommand{\totalPackets}{16,894,368 }
\newcommand{\totalUniqueIPs}{13,050 }
\newcommand{\totalUniqueIPsPlusPL}{26,919 }
\newcommand{\totalPeerLists}{673,078 }
\newcommand{\totalUniqueASs}{1,870}
\newcommand{\totalUniqueASsPlusPL}{2,408}
\newcommand{\totalUniqueIPssupportflag}{1,054 }
\newcommand{\UniqueASNsSupportFlag}{80 }
\newcommand{\SFRpackets}{177,853 }
\newcommand{\totalUniqueIPslastseen}{288 }
\newcommand{\UniqueASNsLastSeen}{1 }
\newcommand{\totalUniqueIPsPLDiv}{554 }
\newcommand{\UniqueASNsPLDiv}{1 }
\newcommand{\totalUniqueIPsPLSim}{553 }
\newcommand{\UniqueASNsPLSim}{1 }
\newcommand{\totalConnections}{181,091 }
\newcommand{\tcpAnomaliesIncHS}{4,387 }
\newcommand{\avgConnDuration}{364.61 }
\newcommand{\medConnDuration}{0.54}
\newcommand{\avgConnCommands}{93.14 }
\newcommand{\medConnCommands}{4}
\newcommand{\shortLivedConnsOne}{85,288 }
\newcommand{\totalUniqueIPsShortTwo}{10 }
\newcommand{\UniqueASNsShortTwo}{7 }
\newcommand{\shortLivedOneMaxIP}{65,340 }
\newcommand{\throttledTSmedianfreq}{61.02 }
\newcommand{\throttledTSConns}{121 }
\newcommand{\uniqueIPsthrottledTS}{27 }
\newcommand{\UniqueASNsthrottledTS}{21 }
\newcommand{\pingFloodingConns}{2,900 }
\newcommand{\pingFloodingIPs}{501 }
\newcommand{\pingFloodingASNs}{1 }
\newcommand{\avgInDegree}{957.66}
\newcommand{\medianInDegree}{8}
\newcommand{\uniqueIPsIDAnom}{635 }
\newcommand{\uniqueASIDAnom}{150 }
\newcommand{\uniqueIDclusterIPs}{265 }
\newcommand{\uniqueIDclusterASs}{33 }
\newcommand{\numIDcluster}{5 }
\newcommand{\uniqueIDIPinterIPs}{880 }
\newcommand{\uniqueIDIPinterASNs}{171 }
\newcommand{\medianSubnetPeers}{1}
\newcommand{\totalSusPeers}{1,877 }
\newcommand{\susAndReachable}{1,666 }
\newcommand{\totalUniqueIPsSign}{1,123 }
\newcommand{\UniqueASNsSign}{5 }
\newcommand{\LionLinkPeers}{1,582 }
\newcommand{\incomingConnSatwithout}{18.05}
\newcommand{\outgoingConnSatwithout}{7.78}
\newcommand{\incomingConnSatAdvwithout}{0.94}
\newcommand{\outgoingConnSatAdvwithout}{6.69}
\newcommand{\totalincomingConnSatAdvwithout}{1.33 }
\newcommand{\totaloutgoingConnSatAdvwithout}{0.74 }
\newcommand{\totalincomingConnSatwithout}{25.60 }
\newcommand{\totaloutgoingConnSatwithout}{0.81 }
\newcommand{\avgPeerPoisoning}{16.22}
\newcommand{\percentageAnomNet}{14.38}
\newcommand{\percentageAnomReach}{12.77}
\newcommand{\totalAnomalous}{1,877 }
\newcommand{\myUniquensPeers}{734 }

\newcommand{\incomingConnSatwith}{15.21}
\newcommand{\outgoingConnSatwith}{3.18}
\newcommand{\incomingConnSatAdvwith}{2.28 }
\newcommand{\outgoingConnSatAdvwith}{2.79}
\newcommand{\totalincomingConnSatAdvwith}{2.39 }
\newcommand{\totaloutgoingConnSatAdvwith}{0.84 }
\newcommand{\totalincomingConnSatwith}{16.04 }
\newcommand{\totaloutgoingConnSatwith}{0.96 }

\begin{abstract}
  Monero, the leading privacy-focused cryptocurrency, relies on a peer-to-peer (P2P) network to propagate transactions and blocks. Growing evidence suggests that non-standard nodes exist in the network, posing as honest nodes but are perhaps intended for monitoring the network and spying on other nodes. However, our understanding of the detection and analysis of anomalous peer behavior remains limited. This paper presents a systematic methodology for detecting anomalous peer behavior in P2P cryptocurrency networks, which we apply in a forensic case study of Monero's network. We collected and analyzed over 240 hours of network traffic captured from five distinct vantage points worldwide. We further present a formal framework which allows us to analytically define and classify anomalous patterns in P2P cryptocurrency networks. Our detection methodology defines detection rules directly applicable to real-time monitoring. Our analysis reveals the presence of non-standard peers in the network where approximately 14.38\% (1,877 of 13,050) of the observed peers exhibit anomalous behavior. Of these, at least 1,582 peers show patterns consistent with control by a single entity.
These peers exhibit behavioral patterns consistent with a Sybil attack coupled with self-reinforcing peer list poisoning, which establishes the preconditions for degrading Monero's privacy guarantees and undermining network decentralization. To support reproducibility and enable network operators to protect themselves, we release our examination pipeline to identify and block suspicious peers based on newly captured network traffic.

\end{abstract}

\section{Introduction} 
The security and privacy of cryptocurrencies critically depends on the underlying P2P networks as adversaries exploiting network-level vulnerabilities can bypass security and privacy mechanisms applied at the protocol and application layer, risking the network's integrity even for privacy-focused cryptocurrencies. Ten years after its release, Monero continues to be the dominant privacy coin. With a 24-hour trading volume exceeding 100 million USD\footnote{https://coinmarketcap.com/currencies/monero/}, Monero demonstrates that users increasingly demand financial privacy comparable to physical cash in traditional monetary systems. Simultaneously, empirical evidence has demonstrated a notable migration from Bitcoin to Monero in illicit online marketplaces~\cite{BAHAMAZAVA2022301377}. 

While general blockchain analysis platforms exist~\cite{kalodner2020blocksci} Monero's privacy techniques have demonstrated robust protection against sophisticated blockchain analysis~\cite{hammad2024monerotraceabilityheuristicswallet,möser2018empiricalanalysistraceabilitymonero}. Consequently, rather than breaking cryptographic primitives, adversaries appear to be increasingly targeting the underlying P2P network layer~\cite{biryukov2019deanonymization,cao2020exploring} to exploit network-level vulnerabilities that undermine claimed privacy guarantees. In 2023, community developers identified a few highly saturated /24 subnets, each hosting close to the full 254-host range of peers, opening many short-lived connections in the Bitcoin network as well as the Monero P2P network, with this pattern continuing into 2024\footnote{https://b10c.me/blog/013-one-year-update-on-linkinglion}. In September of 2024, an independent community analysis\footnote{https://www.digilol.net/blog/chainanalysis-malicious-xmr.html} followed, discussing the potential intrusion of non-standard monitoring nodes within the P2P network. Such monitoring nodes would be positioned to observe RPC calls, potentially linking invoked transactions to real-world IP addresses, a capability that directly undermines Monero's privacy guarantees. 
This capability not only defeats Monero's privacy promise but may also lower the bar for targeted attacks such as focused eclipse attacks against specific users, although eclipse feasibility depends on the details of peer discovery and connection management and several concrete vectors have since been closed. By late 2024, these concerns prompted Monero developers to issue an official warning\footnote{https://github.com/monero-project/meta/issues/1124} and recommend node operators make use of a protective ban list. However, no public explanation was provided of how this ban list was compiled or what criteria were used to identify malicious nodes, creating a significant gap in the understanding of this network-level threat.

A variety of previously studied attacks on blockchain P2P networks~\cite{apostolaki2017hijacking,biryukov2019deanonymization,cao2020exploring,heilman2015eclipse,saad2022partitioningbtc,shi2025eclipse} show that whilst attacks are feasible, an adversary first needs to infiltrate the network with peers frequently diverting from standard configurations. Detection of such non-standard peers remains limited (Sec.~\ref{sec:related}), and our work fills this gap by systematically characterizing these subliminal network intrusions.

\subparagraph{Contributions:}
We investigated Monero's P2P network for about six months with repeated captures of network traffic and analysis cycles. 
This preliminary analysis revealed that detection cost differs sharply across anomaly classes. Some deviations are visible in the first message a peer sends and others only in the timing of a single connection over its first minutes. Others emerge only once compared against one another. Our final measurement design accommodated the slowest of these.
This paper presents a forensic case study, based on a two-day, five-vantage-point traffic capture from June 2025.
To our knowledge, this is the first work to systematically detect passive, in-protocol monitoring nodes in a cryptocurrency P2P network, rather than the disruptive attackers targeted by prior detection systems. Our detection angles are tailored to the cryptocurrency setting and exploit Monero's presumed software homogeneity, under which any systematic deviation becomes a detectable fingerprint.
Our systematic examination framework introduces a novel classification: protocol-level anomalies (Sec.~\ref{subsec:protocolanomalies}) categorized into syntactic violations, identifying evident deviations from protocol specifications; semantic inconsistencies, going beyond structural validation and analyzing logical consistency; and behavioral deviations, including a temporal context separating standard state sequences from anomalous behavior. Furthermore, we advance this peer-based taxonomy by including network structural patterns (Sec.~\ref{subsec:networkstructure}) and node attributes (Sec.~\ref{subsec:nodeattributes}) commonly found in P2P environments. 
Despite Monero's P2P network presuming homogeneous node software deployment across all participating peers, our examination (Sec.~\ref{sec:monerosexamination}) revealed that approximately \percentageAnomNet\% of peers observed during our measurement period exhibited anomalous behavior, with one entity estimated to control at least \LionLinkPeers of the observed nodes. These nodes distort Monero's peer discovery mechanism to gain strategic network positions, enabling an effective first-spy estimator.
Our anomaly detection approach complements community efforts by providing a comprehensive framework with empirical evidence for identifying non-standard nodes, raising concerns about Monero's privacy guarantees and network decentralization.
To ensure reproducibility and to facilitate follow-up work, we release our GitHub repository and dataset, allowing every Monero node operator to replicate our results and analyze their own network traffic, identify unusual peers, and block them if desired. 

\section{Related Works}\label{sec:related}
Prior work on Monero and related P2P layers spans topology inference, transaction deanonymization, and eclipse attacks, alongside analogous vulnerabilities in Bitcoin and Ethereum.

\subparagraph{Monero's Network Attacks.}
Cao et al.~\cite{cao2020exploring} demonstrated a topology learning attack by exploiting \texttt{last\_seen} timestamps in Handshake and Timed Sync responses, achieving 97.98\% recall for inferring node connections across crawled data. In response, developers removed the timestamp~\footnote{https://github.com/monero-project/monero/pull/5682} from Peer Lists (PLs), neutralizing the attack while leaving network centralization unaddressed. Shi et al.~\cite{shi2024deanonymizing} later correlated transactions with IP addresses by injecting malicious Monero Tor hidden service nodes and watermarking Tor circuits to link onion and real addresses. Shi et al.~\cite{shi2025eclipse} then demonstrated the first practical eclipse attack against Monero, combining peer-discovery flooding, priority-list population, and double-spend-induced connection drops to eclipse a victim within minutes using 1,020 IPs. A 2024 update addressed the reset vector~\footnote{https://github.com/monero-project/monero/pull/9218}, though new vectors may emerge.

\subparagraph{Network-Level Attacks on Bitcoin and Ethereum.} Eclipse attacks are well studied here. Heilman et al.~\cite{heilman2015eclipse} monopolized a target's peer list and waited for a restart to achieve isolation; Tran et al.~\cite{tran2020stealthier} introduced Erebus, a stealthier AS-level attack replacing TCP connections without direct occupation; and Marcus et al.~\cite{marcus2018low} forced restarts and flooded incoming connections in a low-resource variant. These works highlight the persistent vulnerability of P2P networks to network-layer isolation.

\subparagraph{Anomaly Detection in P2P Networks.} Sybil attacks were formalized by Douceur~\cite{douceur2022sybil}, who showed that without a central identity authority they are unpreventable in open systems, and have since been measured in deployed P2P networks. Wang and Kangasharju~\cite{wang2012real} identified large-scale Sybil deployments in BitTorrent's Mainline DHT precisely by their partial protocol implementation, and Neudecker and Hartenstein~\cite{neudecker2019network} survey the corresponding network-layer design space for permissionless blockchains. These works detect Sybils through address concentration, active crawling, or honeypots. We instead fingerprint per-packet and per-connection protocol behavior observed passively from peers that connect to us.

\section{Background}\label{sec:background}
Monero differs from generic P2P networks as there is effectively one maintained client. Within its P2P network, Monero aims for homogeneous-looking traffic as core fungibility design goal, any systematic deviation requires deliberate code modification. Throughout this paper, we use three related but distinct terms with increasing evidential burden. We call a node non-standard if it deviates from the Monero specification at any layer, including protocol fields, message timing, or software configuration. We call a node's behavior anomalous when it produces observable patterns that deviate from what a standard node would produce. Finally, we reserve the terms malicious and adversarial for cases where corroborating evidence supports an intentional, hostile interpretation of non-standard behavior.

\begin{figure}[t]
  \begin{subfigure}[t]{0.48\textwidth}
    \begin{tikzpicture}[
    node/.style={rectangle, draw, minimum width=1.6cm, minimum height=0.6cm, align=center},
    arrow/.style={-{Stealth[length=3mm]}, thick},
    message/.style={font=\small, align=left},
    timing/.style={font=\scriptsize, blue}
]

\node[node] (nodeA) at (0,0) {Node A};
\node[node] (nodeB) at (5,0) {Node B};

\draw[arrow] ($(nodeA.south) + (0,-0.2)$) -- ($(nodeB.south) + (0,-0.2)$);
\node[message] at (2.5,-0.3) {Handshake Req};
\node[message, font=\scriptsize] at (2.5,-0.8) {network\_id, peer\_id, my\_port,\\\textcolor{blue}{support\_flags}, payload\_data};
\draw[arrow] ($(nodeB.south) + (0,-1.3)$) -- ($(nodeA.south) + (0,-1.3)$);
\node[message] at (2.5,-1.4) {Handshake Res};
\node[message, font=\scriptsize] at (2.5,-1.9) {network\_id, peer\_id, support\_flags,\\payload\_data, \textcolor{blue}{local\_peerlist\_new}};

\draw[arrow] ($(nodeB.south) + (0,-2.4)$) -- ($(nodeA.south) + (0,-2.4)$);
\node[message] at (2.5,-2.5) {Ping};
\draw[arrow] ($(nodeA.south) + (0,-2.9)$) -- ($(nodeB.south) + (0,-2.9)$);
\node[message] at (2.5,-3.0) {Pong};
\node[message, font=\scriptsize] at (2.5,-3.4) {status, peer\_id};

\draw[arrow] ($(nodeA.south) + (0,-3.7)$) -- ($(nodeB.south) + (0,-3.7)$);
\node[message] at (2.5,-3.8) {Timed Sync Req};
\draw[->] (3.6,-3.6) arc (180:-110:0.2);
\node[timing, black] at (3.8,-3.6) {60s};
\node[message, font=\scriptsize] at (2.5,-4.2) {payload\_data};
\draw[arrow] ($(nodeB.south) + (0,-4.5)$) -- ($(nodeA.south) + (0,-4.5)$);
\node[message] at (2.5,-4.6) {Timed Sync Res};
\node[message, font=\scriptsize] at (2.5,-5) {payload\_data, \textcolor{blue}{local\_peerlist\_new}};

\draw[arrow] ($(nodeB.south) + (0,-5.4)$) -- ($(nodeA.south) + (0,-5.4)$);
\node[message] at (2.5,-5.5) {Timed Sync Req};
\draw[blue, ->] (3.6,-5.3) arc (180:-110:0.2);
\node[timing] at (3.8,-5.3) {60s};
\node[message, font=\scriptsize] at (3,-5.9) {payload\_data};
\draw[arrow] ($(nodeA.south) + (0,-6.2)$) -- ($(nodeB.south) + (0,-6.2)$);
\node[message] at (2.5,-6.3) {Timed Sync Res};
\node[message, font=\scriptsize] at (2.5,-6.7) {payload\_data, local\_peerlist\_new};

\draw[dashed, gray] (nodeA.south) -- ($(nodeA.south) + (0,-6.6)$);
\draw[dashed, gray] (nodeB.south) -- ($(nodeB.south) + (0,-6.6)$);
\end{tikzpicture}
    \caption{\textbf{Monero's Default Message Sequence} The figure shows a standard Monero message sequence where node A opens an outgoing connection to node B. The Handshake request initiates the connection. The Ping/Pong sequence verifies that A allows incoming connections and adds A to B's white list. The Timed Sync commands are individual loops to retrieve updated information about the network's topology.}
    \label{fig:moneroMsgSequence}
  \end{subfigure}
  \hfill
  \begin{subfigure}[t]{0.48\textwidth}
    \includegraphics[width=\textwidth]{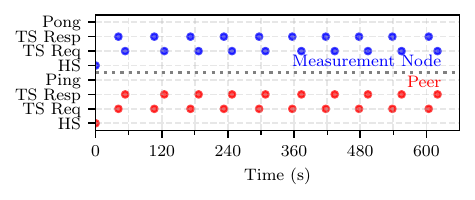}
    \caption{\textbf{Standard Connection.} The graph displays a real sample of a standard incoming connection with each peer sending Timed Sync requests approximately every 60 seconds.}
    \label{fig:standard_connection}
  \end{subfigure}
  \caption{The figures depict how a typical connection in Monero's P2P network should look like, with the default message sequence outlined on the left and a sampled peer connection on the right.}
  \label{fig:typical_connections}
\end{figure}

In the following, we will present an overview of relevant functionalities inside Monero's P2P network. Monero's P2P layer uses a custom application protocol known as the Levin protocol~\cite{monero2024github}, most importantly consisting of a Handshake sequence and a continuous peer discovery mechanism using Timed Sync messages as visualized in Figure~\ref{fig:moneroMsgSequence} and seen for a real sample peer in Figure~\ref{fig:standard_connection}. 

\subparagraph{Handshake Sequence.}
To initialize a connection, node A sends a Handshake request to node B, sharing a unique network ID, node ID, and support flags. Since 2018, the support flag is 1 by default\footnote{https://github.com/monero-project/monero/pull/3191}, indicating the usage of fluffy blocks, where nodes send compact block representations that contain mostly transaction IDs rather than full transaction data.
If B has capacity, B responds with its own node and payload data and additionally a PL containing a maximum of 250 peers as part of Monero's peer discovery mechanism.

\subparagraph{Peer Discovery Mechanism.}
Monero's peer discovery combines local PL management and PL exchanges between neighbors. An initial PL exchange occurs as part of the Handshake, after which both peers enter a 60-second loop that schedules Timed Sync messages to request and send PLs. 
Each Timed Sync response contains up to 250 white list entries, avoiding duplicates during a continuous exchange. Peer entries include IP address, port, network ID, and optionally pruning seed, RPC port, and RPC cost in credits per hash.
Local PL management employs a white list for up to 1,000 validated peers and a gray list for up to 5,000 unvalidated peers. Received peers are initially added to the gray list. Every 60 seconds, the node randomly selects a gray list peer, tests reachability via Ping, and adds it to the white list upon validation. Similarly, incoming connections advertising an open port are tested via Ping and added to the white list if validated.

\subparagraph{Offense Tracker.}
Monero applies a penalty mechanism to score peer behavior and drop connections when necessary~\cite{monero2024github}. However, the mechanism is limited and evaluates specific behaviors in the P2P communication like blockchain height discrepancies, transaction validity, or block verification failures. Each peer maintains a score that decreases when problematic behaviors are detected, such as sending invalid blocks, providing inconsistent blockchain data, or attempting double-spend transactions. When a neighboring peer's score falls below a critical threshold, the connection to this peer is terminated and it may be temporarily banned from reconnecting.

\subparagraph{Transaction Propagation.}
Monero's transaction propagation is based on Dandelion++~\cite{fanti2018dandelion++}. Dandelion++ employs two phases to enhance network-level anonymity. In the stem phase, transactions are forwarded to one of two outgoing relay nodes that are periodically switched. Each intermediate node either continues forwarding as stem or transitions to the fluff phase where transactions are broadcast using standard diffusion.
While the protocol provides improved privacy over naive flooding protocols, Sharma et al.~\cite{sharma2022anonymity} demonstrate significant weaknesses against sophisticated adversaries. Using a combination of coordinated graph learning and intersection attacks, even modest adversaries occupying 20\% of the network can significantly reduce the set of potential originators~\cite{sharma2022anonymity}, revealing a substantial gap between formal guarantees and practical security.

\section{Examination Framework \label{sec:examinationframework}}
This section introduces our methodology for identifying and categorizing anomalous patterns. The framework provides a comprehensive analytical approach that extends beyond simple protocol compliance checking to encompass content, behavioral, structural, and attribute-based detection methods that can be easily applied to a variety of P2P networks. We establish formal definitions for the core components of our anomaly examination. Let $G = (V, E)$ represent a P2P network where $V$ is the set of peers and $E \subseteq V \times V$ is the set of (logically) bidirectional connections. Each connection $e \in E$ carries a time-ordered sequence of messages $\mathcal{M}_{e} = \langle m_1, m_2, \ldots, m_k \rangle$, where each message has an associated type from the set $\mathcal{T}$ of all protocol message types.

\subsection{Protocol-Level Anomaly Detection \label{subsec:protocolanomalies}}
Protocol-level anomalies are observable behavioral deviations that indicate non-standard implementations or configurations adapted from traditional signature-based and specification-based detection methodologies for distributed network environments. We define three distinct categories based on detection complexity and required domain knowledge.

\subsubsection{Syntactic Violations}\label{subsubsec:synviol} To identify syntactic violations, we systematically extract and validate all available protocol fields against current specifications, diverging from traditional signature-based approaches that target specific known attack patterns. Rather than searching for predetermined malicious signatures, our heuristic approach performs exhaustive field-level validation to identify any deviation from syntactic correctness. This includes:
\begin{itemize}
    \item \textbf{Missing Required Fields}: Absence of mandatory protocol elements as defined in current specifications
    \item \textbf{Unexpected Field Presence}: Inclusion of deprecated or non-standard protocol elements  
    \item \textbf{Malformed Message Structures}: Violations of defined message syntax and encoding standards
\end{itemize}
Through exhaustive field-level compliance checking against canonical protocol definitions, this approach identifies nodes running non-standard protocol implementations. While efficient, this methodology cannot detect more sophisticated anomalies or attacks that operate within syntactic correctness boundaries.

\subparagraph{Syntactic Formalization.}
For each type $t \in \mathcal{T}$, let $\mathcal{F}_{req}(t)$ and $\mathcal{F}_{opt}(t)$ represent message type-specific required and optional fields. 
For a message $m$ of type $t$, let $\text{dom}(\phi_m)$ denote the set of fields present in $m$. The message is syntactically anomalous, if:
\begin{equation}
\begin{gathered}
    \text{SynAnom}(m) = \mathcal{F}_{req}(t) \not\subseteq \text{dom}(\phi_m) \lor \text{dom}(\phi_m) \not\subseteq (\mathcal{F}_{req}(t) \cup \mathcal{F}_{opt}(t))
\end{gathered}
\end{equation}
\subsubsection{Content Anomalies}\label{subsubsec:contanom} Content Anomalies extend beyond structural validation to analyze logical content consistency. These subtle anomalies involve correctly formatted messages containing logically inconsistent or fabricated information that requires domain-specific knowledge for detection.
Detection methodologies include:
\begin{itemize}
\item \textbf{Baseline Analysis}: Comparison of observed pattern with established expected content patterns
\item \textbf{Cross-Peer Comparison}: Comparing information consistency across multiple sources
\end{itemize}
This approach effectively identifies content-based attacks that appear as legitimate protocol exchanges, particularly useful for detecting coordinated misinformation campaigns across multiple peers.

\subparagraph{Content Formalization.} For content anomalies we model validation through baseline comparison and cross-peer comparison. For a message $m$ of type $t$ with content $\text{c}_m$, the message is anomalous if:
\begin{equation}
\begin{gathered}
    \text{ContAnom}(m) = \delta_t(\text{c}_m, \mathcal{B}_{t}) \geq \tau_{t} 
    \lor 
    \rho_t(\{m\} \cup \mathcal{M}_{cross,t}^{(m)}) \notin [\tau_{c,min}^{t}, \tau_{c,max}^{t}] 
\end{gathered}
\end{equation}
where $\delta_t$ measures content deviation from the baseline against threshold $\tau_{t}$ for message type $t$. $\rho_t$ computes cross-peer comparison scores for message set $\mathcal{M}_{cross,t}^{(m)}$, containing all messages from distinct peers of the same type. $[\tau_{c,min}^{t}, \tau_{c,max}^{t}] \subseteq [0,1]$ defines the expected scoring range for type $t$. 

\subsubsection{Behavioral Pattern Deviations}\label{subsubsec:behavanom} Behavioral pattern deviations extend beyond static protocol validation to examine dynamic communication patterns, capturing anomalies that manifest through temporal and sequential deviations rather than explicit specification violations. It follows stateful protocol analysis principles~\cite{LIAO201316}, analyzing communication timing, frequency, and sequence patterns. These anomalies manifest through deviations from expected interaction patterns rather than explicit protocol violations.
Key behavioral indicators include:
\begin{itemize}
\item \textbf{Timing and Frequency Anomalies}: Deviations from expected message timing and frequencies
\item \textbf{Sequence Violations}: Non-standard state transitions
\item \textbf{Connection Pattern Anomalies}: Unusual connection establishment and termination behaviors
\end{itemize}
Our methodology models expected protocol behavior, establishing definitive and statistical baseline patterns that enable detection of subtle deviations indicating reconnaissance activities, resource exhaustion attacks, or protocol manipulation attempts. This approach captures sophisticated attacks that operate within protocol specifications but exhibit anomalous temporal or sequential characteristics.

\subparagraph{Behavioral Formalization.} For behavioral anomalies we analyze communication patterns across connections and peers. For a connection $e \in E$, behavioral anomalies are detected through:
\begin{equation}
\begin{gathered}
    \text{BehavAnom}(e, v) = \\
    \text{sequence}(\mathcal{M}_e) \notin \mathcal{S}_{standard} 
    \lor \delta(E^v, \mathcal{B}_{conn}) > \tau_{conn} 
    \lor |\text{timings}(\mathcal{M}_e^t) - \text{timing}_{standard}^t| > \epsilon_t 
\end{gathered}
\end{equation}
where $\text{sequence}(\mathcal{M}_e)$ extracts the sequence of message types from connection $e$, $\mathcal{S}_{standard}$ represents standard sequences, $\text{timings}(\mathcal{M}_e^t)$ extracts the timing frequency of messages of type $t$ in connection $e$, $\text{timing}_{standard}^t$ represents standard timing for message type $t$, $\epsilon$ is the deviation tolerance threshold for message type $t$, $\mathcal{B}_{conn}$ is the baseline for connection patterns, $E^v$ denotes all connections by peer $v$, and $\delta$ measures deviation with threshold $\tau_{conn}$.

\subsection{Network Structure \label{subsec:networkstructure}}
We extend our examination by incorporating classic graph-theoretic concepts, including: 
\begin{itemize}
    \item \textbf{Centrality Anomalies}: Detecting nodes with unusual connectivity patterns 
    \item \textbf{Community Detection}: Identifying unusual clustering patterns indicating coordinated malicious activity
\end{itemize}
P2P networks like cryptocurrency networks rely on decentralization and aim to fully decentralize the network. A certain degree of centralization can naturally occur due to geographically more active locations. Nonetheless, prior research by Zabla et al. and Beikverdi et al. reveal concerning trends toward increased centralization in major blockchain networks~\cite{beikverdi2015trendcentralization,ZABKA2024102696}.
In systems with topology obfuscation such as Monero, the true connection graph is not observable. We therefore apply these concepts to an observable proxy, the promotion topology defined in Sec.~\ref{subsec:promtop}. We instantiate centrality as promotional influence but leave out community detection, as promotion edges would yield communities of advertisement rather than of connectivity.

\subsection{Node Attributes \label{subsec:nodeattributes}}
At last, we introduce a form of attribute fingerprinting based on various available attributes such as IP addresses, unique node identifier, RPC services, or pruning seeds for applicable cryptocurrencies. We include in our analysis: 
\begin{itemize}
    \item \textbf{Identity Correlation}: Correlating identities across the network 
    \item \textbf{Geographic and ASN Analysis}: Analysis of infrastructure distribution patterns
\end{itemize}
Unlike protocol-level anomalies, network structure and node attribute assist our analysis in finding potential clusters by adding a network-wide perspective. We thus omit formal mathematical definitions.
We now instantiate this framework on data collected from Monero's live network.

\section{Case Study: Monero's P2P Network 
\label{sec:monerosexamination}}
We implement a comprehensive detection methodology to identify and quantify anomalous behavior within Monero's P2P network, applying the framework established in Sec.~\ref{sec:examinationframework}. 
The methodology leverages network traffic captured from unmodified P2P nodes participating in the presumably homogeneous network as standard pruned nodes with both incoming and outgoing connections. 
Our investigation spanned several months with repeated informal snapshots. In these preliminary observations, the same classes of anomalies recurred across capture cycles, which motivated a final measurement design of two consecutive 24-hour snapshots capturing this behavior at scale while keeping storage and computation feasible.

\subsection{Data Collection}
Our dataset constitutes a snapshot: five unmodified vantage points, two configurations each, captured over two days in June 2025. It is therefore a forensic record of the network state at that time, not a longitudinal characterization. All percentages reported in this section refer to peers observed during this window. Since anomalous infrastructure can be redeployed or withdrawn, current prevalence may differ. However, the released pipeline enables operators to repeat the measurement on newly captured traffic.
\begin{itemize}
    \item \textbf{Locations}: San Francisco, Amsterdam, Bangalore, Singapore, and Sydney
    \item \textbf{Default Configuration}: 12 outgoing connections and no ban list enabled
    \item \textbf{Community Configuration}: 32 outgoing connections and a pre-loaded ban list\footnote{https://github.com/monero-project/meta/issues/1124}
    \item \textbf{Timeframe}: Network traffic captured June 2–4, 2025, with each configuration running for 24 hours
\end{itemize}
Following data collection, network packet analysis was conducted using Wireshark~\cite{wireshark447} to extract protocol information from identified Monero network traffic. We collected all available data fields, including command identifiers, node attributes, and payload data required for PL analysis. Each PL was systematically processed to extract complete peer entries with all associated metadata fields. This approach yielded a timestamped dataset enabling granular analysis of individual network connections, peer behaviors, and network representations.

\subsection{Examination Results}
In aggregate, the measurement nodes captured a total of 240 hours of network traffic. The collected dataset statistics are reported in Table~\ref{tab:node-results} across all measurement nodes.
\begin{table}[t]
  \caption{\textbf{Individual and Aggregated Node Results.} The table shows the amount of total Monero packets sent and received. The IPs are represented as IPs we have had a direct connection with and additionally total IPs we have seen inside the collected PLs.}
  \label{tab:node-results}
  \centering
  \begin{tabular}{p{2.3cm}|p{1.4cm}|p{1.4cm}|p{1.4cm}|p{1.4cm}|p{1.4cm}|p{1.5cm}}
    & SF & AMS & BLR & SG & SYD & Total \\
    \hline
    \hline
    Monero Packets & 3,573,738 & 4,807,042 & 1,902,680 & 3,004,744 & 3,606,164 & \totalPackets \\
    \hline
    PLs received & 135,242 & 171,956 & 113,753 & 118,692 & 133,435 & \totalPeerLists \\
    \hline
    IPs connected (ASs) & 6,573 (1,246) & 6,697 (1,229) & 6,226 (1,170) & 6,396 (1,173) & 6,507 (1,194) & \totalUniqueIPs (\totalUniqueASs) \\
    \hline
    IPs seen (ASs) & 14,401 (1,668) & 17,784 (1,833) & 15,191 (1,694) & 14,237 (1,630) & 15,935 (1,695) & \totalUniqueIPsPlusPL (2,480)\\
  \end{tabular}
\end{table}


The extracted PLs contained multiple IPs that are either stale or didn't have a connection established with any of our measurement nodes. These IPs yield information on subnet and AS affiliation but cannot be classified further. 
The examination results are aggregated over the five nodes.
Unless noted otherwise, all examination results derive from the default-configuration nodes, some community-configuration results are reported in Sec.~\ref{subsec:defense}.
All data\footnote{https://zenodo.org/records/16947083} and the complete examination pipeline\footnote{https://anonymous.4open.science/r/monero-traffic-analysis-2528} is publicly available to reproduce the results in this paper. 

We present the results in order of increasing evidential weight rather than framework order, to build a chain from structural and content indicators toward behavioral and syntactic fingerprinting.

\subsubsection{IP Address Affiliation \label{subusbsec:ipaffiliation}}
We begin by characterizing the global distribution of IP addresses, as part of the network structure (Sec.~\ref{subsec:networkstructure}), across both direct connections and PLs, establishing a structural baseline against which anomalies are measured. 
As noted, the PLs include IP addresses with varying temporal validity, as PLs may contain stale entries from disconnected nodes or outdated IPs. This limits what PL-derived counts can support. We therefore use PL data only for structural claims about subnet and AS concentration, which are robust to individual entries being outdated, and never as a count of concurrently online peers.
Our analysis reveals two concerning findings.
\begin{figure}[t]
\begin{subfigure}[c]{0.48\textwidth}
    \includegraphics[width=1\linewidth]{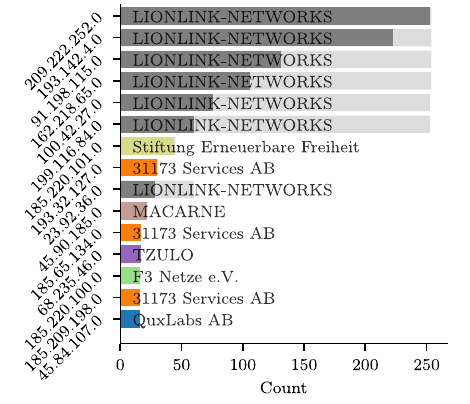}
    \caption{\textbf{AS Subnet Distribution Of Established Connections.} Each bar represents a /24 subnet with colored ASs. The light gray bars show IPs retrieved from PLs without an established connection, resulting in a few fully saturated /24 subnets and multiple saturated subnets are affiliated under the same AS organization.}
    \label{fig:asn_subnet_distribution}
  \end{subfigure}
  \hfill
  \begin{subfigure}[c]{0.48\textwidth}
    \centering
    \pgfkeys{%
    /calloutquote/.cd,
    width/.code                   =  {\def\calloutquotewidth{#1}},
    position/.code                =  {\def\calloutquotepos{#1}}, 
    author/.code                  =  {\def\calloutquoteauthor{#1}},
    /calloutquote/.unknown/.code   =  {\let\searchname=\pgfkeyscurrentname
                                 \pgfkeysalso{\searchname/.try=#1,                                
    /tikz/\searchname/.retry=#1},\pgfkeysalso{\searchname/.try=#1,
                                  /pgf/\searchname/.retry=#1}}
                            }  
\newcommand\calloutquote[3][]{%
       \pgfkeys{/calloutquote/.cd,
         width               = 5cm,
         position            = {(0,-1)},
         author              = {}}
  \pgfqkeys{/calloutquote}{#1}                   
  \node [rectangle callout,callout relative pointer={\calloutquotepos},text width=\calloutquotewidth,/calloutquote/.cd,
     #1] (tmpcall) at #2 {#3};
  \node at (tmpcall.pointer){\calloutquoteauthor};    
}  

\begin{tikzpicture}[
    scale=0.65,
    node/.style={circle, draw, minimum width=0.8cm, minimum height=0.5cm, align=center, font=\scriptsize},
    listheader/.style={rectangle, draw, minimum width=1.0cm, minimum height=0.5cm, align=center, font=\small},
    arrow/.style={-{Stealth[length=2.5mm]}, thick},
    arrowboth/.style={{Stealth[length=2.5mm]}-{Stealth[length=2.5mm]}, thick},
    message/.style={font=\scriptsize, align=left},
]

\node[font=\small, align=center] at (1.2,-3.5) {\textbf{Temporal Anomaly}};

\node[node] (nodeA) at (0,3.0) {IP\\1.2.3.4};
\node[node] (nodeB) at (0,0.5) {IP\\1.2.3.4};
\node[node] (nodeC) at (0,-2.0) {IP\\1.2.3.4};

\calloutquote[author=,width=1cm,position={(-0.8,-0.2)},fill=green!30,rounded corners,font=\scriptsize]{(2.6,3.5)}{I am abcd}
\calloutquote[author=,width=1cm,position={(-0.8,-0.2)},fill=green!30,rounded corners,font=\scriptsize]{(2.6,1.0)}{I am efgh}
\calloutquote[author=,width=1cm,position={(-0.8,-0.2)},fill=red!30,rounded corners,font=\scriptsize]{(2.6,-1.5)}{I am abcd\\again}

\draw[arrow] (nodeA.south) -- (nodeB.north) node[midway, right, font=\scriptsize] {reboot};
\draw[arrow] (nodeB.south) -- (nodeC.north) node[midway, right, font=\scriptsize] {reboot};

\node[font=\small, align=center] at (6.35,-3.5) {\textbf{IP:ID Clustering}};

\node[listheader] (IP_Addresses) at (5.1,3.5) {IPs};          
\node[listheader] (Peer_IDs)     at (7.6,3.5) {Peer IDs};

\node[node] (Peer1) at (5.1,2.2) {1.2.3.4};
\node[node] (ID1)   at (7.6,2.2) {abcdef};

\node[node] (Peer2) at (5.1,0.4) {2.2.3.4};
\node[node] (ID2)   at (7.6,0.4) {ghijkl};

\node at (5.1,-0.6) {\textbf{$\vdots$}};
\node at (7.6,-0.6) {\textbf{$\vdots$}};

\node[node] (PeerN) at (5.1,-2.0) {a.b.c.d};
\node[node] (IDN)   at (7.6,-2.0) {NNNN};

\draw[arrowboth] (Peer1) -- (ID1);
\draw[arrowboth] (Peer1) -- (IDN);
\draw[arrowboth] (Peer2) -- (ID2);
\draw[arrowboth] (Peer2) -- (IDN);
\draw[arrowboth] (PeerN) -- (ID1);
\draw[arrowboth] (PeerN) -- (IDN);

\end{tikzpicture}
    \caption{\textbf{Peer ID Anomalies.} Monero's unique peer identifier is randomly generated during the start of a node, keeps static during the runtime, and newly generated again between every reboot. As the ID is not stored or loaded, after a node rebooted and identifies with a new ID, it should not have access to the older ones. Simultaneously, IPs cluster themselves along shared IDs, building not fully interconnected clusters of IP addresses aggregating behind multiple distinct identifiers.}
    \label{fig:nodeIDanom}
  \end{subfigure}
  \caption{The figures show anomaly patterns in Monero's IP distribution (left side) and when matched to the peer identifier (right side).}
\end{figure}
\subparagraph{Subnet Saturation.} We observed six /24 subnets fully saturated with 254 promoted IP addresses each. Additionally, Figure~\ref{fig:asn_subnet_distribution} shows that multiple subnets exhibit a population of ten or more peers.

\subparagraph{AS Concentration.} All six saturated subnets are assigned to a single AS, which contains an additional densely populated subnet, totaling \LionLinkPeers peers within AS LIONLINK-NETWORKS (ASLN), occupying a large fraction of the peers observed in our measurement. 

\subparagraph{} A high number of peers inside a few ASs can be expected and follows an ongoing centralization trend in decentralized systems~\cite{cao2020exploring,polinski2025peertube,saad2022partitioningbtc}. However, a high number of peers inside the same subnet is a common indicator that these peers share a single operator. Figure~\ref{fig:asn_subnet_distribution} shows the most represented subnets along with their respective AS organization. Whereas the median saturation of any subnet is at \medianSubnetPeers, the top ones are fully saturated, and the most saturated ones share the ASLN.

\subsubsection{Monero's Peer Identifier \label{subusbsec:monID}}
We examined the nodes' peer identifier as one of Monero's node attributes (Sec.~\ref{subsec:nodeattributes}). Peer identifiers are unique IDs randomly created during the start of the node. If a node is rebooted a fresh ID is created and if it reboots once or multiple times while the IP persists, an IP address naturally announces two or more identifiers. Similarly, if the peer's network IP changes due to ISP mapping or temporarily connecting to a VPN, it will keep its identifier while being associated with multiple IPs. However, by adding a temporal constraint, we compared the identifiers to the complete set of peers to identify patterns of back and forth changing peer IDs and clusters of IP addresses aggregating behind multiple distinct identifiers. This creates an anomaly in the network's node attributes as visualized in Figure~\ref{fig:nodeIDanom}.
We employed two analytical approaches to identify the respective patterns.

\subparagraph{Temporal ID Anomaly.} First, we sorted all collected packets that contain an identifier and checked the temporal constraint for ID announcements. An anomaly of back and forth changing IDs was observed for:
\uniqueIPsIDAnom unique IPs, across
\uniqueASIDAnom unique ASs. 
\subparagraph{ID:IP Clustering.} As a second approach, we extracted IP addresses that announce multiple distinct IDs and reviewed if such IDs are also used by other IPs to identify clusters where distinct IDs are used by distinct IPs and vice versa as visualized at the right side of Figure~\ref{fig:nodeIDanom}. 
We observed:
\uniqueIDclusterIPs unique IPs clustered in \numIDcluster distinct clusters, across
\uniqueIDclusterASs unique ASs.

\subparagraph{}We observed a variety of ASs showing this anomaly and we acknowledge that peers might potentially aggregate behind a shared VPN endpoint. However, comparing it to the rest of the network, it does create an anomaly as 88\% of the connected IP addresses announce only a single identifier. Moreover, the Monero Research Lab uses an active probing technique to identify peers switching IDs\footnote{https://github.com/Rucknium/xmrnetscan/tree/main}, and adding affected IPs to their published ban list. Finally, one of the five identified clusters contains 109 unique IPs exclusively within the densely populated ASLN.

\subsubsection{PL Anomalies \label{subsubsec:planom}}
The peer discovery mechanism is central to Monero's network health, independent of any blockchain or consensus logic, it governs how nodes locate and connect to peers. 
Here, we applied ContAnom (Sec.~\ref{subsubsec:contanom}) to examine PL content with two approaches. First, we analyzed raw PL data at the individual IP address level. Second, we aggregated IP addresses by /24 CIDR subnet to identify potential clustering patterns. We selected /24 subnets as our aggregation unit because nodes within the same /24 subnet are likely to be controlled by a single entity, making subnet-level analysis a meaningful clustering indicator. 
Based on the different aggregation methods, we evaluated PLs individually by measuring their diversity and also performed cross-peer comparisons between distinct peers to identify cluster of peers sharing similarities. For each approach, we included only lists containing at least 225 entries (allowing a 10\% variation) to establish a comparable baseline of similarly sized lists.

\begin{figure}[t]
  \begin{subfigure}[c]{0.48\textwidth}
    \includegraphics[width=1\linewidth]{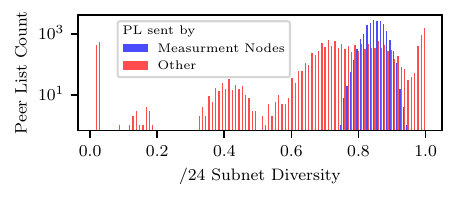}
    \caption{\textbf{PL Diversity.} Distribution of PL diversity scores of individual PLs. Blue bars show PLs sent by our default measurement nodes, red bars represent PLs sent by all other IPs. Diversity is the ratio of unique /24 subnets to total peers. A group of lists accumulates at the left side of the figure, representing less than seven subnets among 250 peers.}
    \label{fig:pl_diversities}
  \end{subfigure}
  \hfill
  \begin{subfigure}[c]{0.48\textwidth}
    \includegraphics[width=\linewidth]{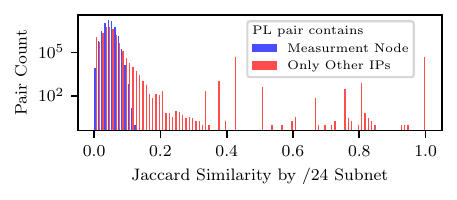}
    \caption{\textbf{PL Similarity.} Distribution of pairwise similarities between all distinct PL pairs excluding same peer comparisons. IPs in the PLs are reduced to its /24 subnet. Blue bars represent comparisons where at least one PL is sent by a measurement node. The distribution shows noticeable exceptions with high similarity scores and hundreds of pairs reaching a perfect overlap are on the right side of the figure.}
    \label{fig:plpairsimilarities}
  \end{subfigure}
  \caption{\textbf{PL Inconsistencies.} The figures display PL inconsistencies. The extreme outliers on the left side for the PL diversities and on the right side for PL similarities represent nodes that are part of the ASLN.}
  \label{fig:pl_anomalies}
\end{figure}

\subparagraph{PL Diversity.} We define diversity as the ratio of unique /24 subnets to total peers in a list. Under standard operation, random selection from the local white list is likely to produce PLs with high subnet diversity. A perfectly diverse list achieves a diversity index of 1.0, while lower values indicate subnet clustering. Modest clustering may occur naturally due to random selection from popular cloud providers, but extreme clustering represents a content anomaly that deviates from expectations of the protocol's random selection specification.

 
Figure~\ref{fig:pl_diversities} shows the distribution of diversity scores for PLs sent by our measurement nodes (blue) and all other IPs (red). All of our measurement nodes consistently achieve diversity scores above 0.7, which aligns with 90\% of all PLs. However, two notable outliers (left side) exhibit diversity scores below 0.028. These lists advertise up to 250 peers distributed across just six or fewer subnets. 
Empirically, we observed \totalUniqueIPsPLDiv unique IPs show this anomalous PL diversity, all residing within the ASLN.

\subparagraph{PL Similarity.} We continued our analysis by measuring a pairwise similarity between PLs with IPs aggregated to its subnets.
We computed standard Jaccard similarity coefficients, which capture the proportion of shared elements between two sets, to quantify the overlap between peer lists from independent IP sources.
For a decentralized P2P network, PLs from independent peers should exhibit low similarity due to the randomized peer selection from distinct local white lists. However, our analysis revealed node pairs with suspiciously high similarity coefficients, including instances of over 90\% and perfect overlap in subnet reduced representations.
Figure~\ref{fig:plpairsimilarities} shows the distribution of pairwise similarities. With several peaks, especially, the peak showing a complete overlap in the subnet representation appears critical. Contrary, for similarity pairs where at least one of the PLs belongs to one of our measurement nodes, the similarity score never exceeds a score above 0.2.

Empirically, \totalUniqueIPsPLSim unique IPs achieve perfect subnet-level overlap in their peer lists, all residing within the ASLN. These peers exclusively advertise subnets in which they themselves reside, forming a promotional network spanning seven subnets. A peer that had itself been poisoned would predominantly advertise a foreign subnet. Self-referential advertisement, where a peer's list is dominated by the subnet it resides in, is therefore more consistent with being the source of the promotion than its recipient.

\subparagraph{Monero's Promotion Topology.}\label{subsec:promtop}
The PL anomalies identified above characterize individual peer behavior. We now examine how these behaviors manifest at the network level. Since Monero implements topology obfuscation to protect routing privacy, we cannot reconstruct the true connection graph. Rather than developing novel topology learning attacks, we examine Monero's ``promotion topology'', revealing structural anomalies in Monero's peer discovery mechanism. We define the promotion topology as a directed graph with IP addresses as nodes. For each received PL, we construct directed edges from the source of the PL to every peer it has promoted within this PL. While this graph does not reflect the network's actual topology at any given time, it captures promotional behaviors. Therefore, multiple edges from one node to another are allowed as multiple promotions naturally occur for extended timeframes. This view has limits: promotion edges reflect advertisement behavior, not established connections, and are biased toward peers that our vantage points happened to observe. It therefore supports claims about promotional prominence, not about actual link-level topology or global node degree.

We observed, that the in-degree distribution of the promotion topology is highly skewed, with a small number of peers receiving a disproportionate share of promotions, consistent with scale-free degree distributions observed in Internet topology~\cite{faloutsos1999power}. At the IP level, individual ASLN addresses do not rank among the highest in-degree nodes. However, when promotions are aggregated by AS and within our captured PLs, ASLN dominates, receiving nearly three times as many promotions as the next largest AS (Hetzner Online GmbH) and more than hundred times as many as the top individual IP. 

\subsubsection{Behavioral Anomalies \label{subsubsec:behaveanom}}
Structural and content anomalies established the presence of non-standard peers. We now apply \emph{BehavAnom} (Sec.~\ref{subsubsec:behavanom}) to examine behavioral deviations that further characterize these nodes. We measured timing, frequency, and sequential deviations across the P2P protocol. Monero's message set comprises base commands that follow deterministic patterns and event commands that exhibit unpredictable timing characteristics. A standard connection sequence consists of two mandatory message exchanges, bidirectional Handshake and periodic Timed Sync commands, each in request-response format. Additionally, when a peer advertises an accessible port for incoming connections, Ping messages are transmitted by the connecting peer to verify the advertised inbound connectivity capability. The Timed Sync exchanges occur at a defined frequency of 60 seconds, initiated by both peers in parallel. 
Figure~\ref{fig:typical_connections} demonstrates the default and baseline behavior. The patterns represent the expected behavior between two standard Monero clients. To identify anomalous behavior, we focus on the three predictable command patterns.
We adopt a connection grouping methodology, filtering our dataset by individual connections before conducting detailed analysis. Connections exhibiting TCP-level anomalies, likely attributable to Wireshark dissector limitations, were excluded from analysis due to their negligible frequency and potential for introducing analytical artifacts.
We located:
\begin{itemize}
    \item \totalConnections total TCP connections. 
    \item Connections last on average \avgConnDuration (median: \medConnDuration) seconds, and
    \item connections contain on average \avgConnCommands (median: \medConnCommands) commands.
\end{itemize}
The discrepancy between average and median is the result of many short-lived connections usually lasting less than one second. 

\subparagraph{Short-lived Connections.} While brief connections may result from legitimate network conditions or operational factors, it becomes critical to determine whether specific peer populations exhibit disproportionate representation in short-lived connection occurrences. We therefore analyzed the distribution of IPs across short-lived connection anomalies, as repeating IPs indicate systematic scanning or probing activities. Empirically, we observed \shortLivedConnsOne connections lasting less than 1 second. Of these, \totalUniqueIPsShortTwo unique IPs have more than ten short-lived connections, across \UniqueASNsShortTwo unique ASs.

All of the observed IP connections completed the Handshake sequence and, in some cases, sent a single Ping or Timed Sync message before terminating. Notably, a single IP address accounts for \shortLivedOneMaxIP connections terminating in under one second. This pattern indicates systematic probing rather than transient network churn. We note that research and community crawlers produce a similar signature and this evidence alone does not distinguish scanning conducted for measurement from scanning conducted for reconnaissance.

\begin{figure}[t]
  \begin{subfigure}[t]{0.48\textwidth}
    \includegraphics[width=1\linewidth]{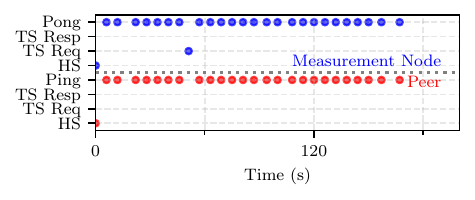}
    \caption{\textbf{Ping Flooding.} A peer spamming Ping messages to our measurement node. After a successful Handshake, the peer spams Ping messages. The measurement node does not crash, treats the connection as usual, and starts at around 60 seconds with the first Timed Sync request. The connection is dropped as the peer does not respond to the request within 120 seconds.}
    \label{fig:ping_flooding}
  \end{subfigure}
  \hspace{0.1cm}
  \begin{subfigure}[t]{0.48\textwidth}
    \includegraphics[width=\linewidth]{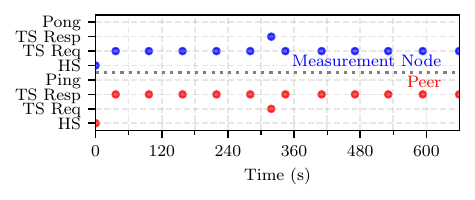}
    \caption{\textbf{Throttled Timed Sync.} Sample peer throttling its Timed Sync request messages. Contrasting to the measurement node using the fixed 60-second frequency for sending requests, the peer follows a non-standard frequency. With a single request within ten minutes while responding timely, the peer extremely deviates from the protocol standard.}
    \label{fig:throttled_ts}
  \end{subfigure}
  \caption{\textbf{Behavioral Anomalies.} Both figures show connection-level deviations from Monero's protocol standard, as observed on our measurement nodes.}
\end{figure}
  
\subparagraph{Ping Flooding.} We observed excessive Ping message transmissions beyond what normal peer verification requires. Figure~\ref{fig:ping_flooding} visualizes a sample connection exhibiting this bursty behavior. Within 24 hours, our measurement nodes received \pingFloodingConns Ping flooding incidents, each delivering a burst of around 23 pings, from \pingFloodingIPs distinct IPs, all residing in the ASLN. Notably, incoming connections should only respond to locally initiated Ping messages to prove their capability to as well accept incoming connections, showing a clear deviation from defined protocol sequences. These connections serve no other apparent protocol function, as evidenced by their uniform disregard for Timed Sync requests. They consistently ignore the first Timed Sync request, causing the connection to drop after the 120-second inactivity timeout, indicating that these nodes neither seek nor sustain normal protocol participation.

\subparagraph{Throttled Timed Sync.} Given the fixed protocol specification requiring Timed Sync requests every 60 seconds, we measured the frequency distribution of these messages across all connections. Figure~\ref{fig:throttled_ts} demonstrates a peer transmitting Timed Sync requests far off the specification.
We initially developed a baseline by analyzing the network's overall distribution of Timed Sync frequencies as some timing variations may be attributed to network latency or node computational load. We average the Timed Sync frequency on each connection and the median is at around \throttledTSmedianfreq seconds, matching with the protocol specifications. However, we identified connections that consistently deviate from the defined 60-second interval.
We set our threshold at 120 seconds, what corresponds to monerod's timeout period after which a node connection drops due to peer unresponsiveness. Any peer with a Timed Sync frequency above 120 seconds must be responsive in general communication. In all cases where the Timed Sync frequency exceeds 120 seconds, peers respond to our requests without notable delay. While we cannot attribute throttled Timed Sync to a single specific attack vector, the behavior requires deliberate source code modification.
Empirically, we observed \throttledTSConns connections have a frequency above 120 and \uniqueIPsthrottledTS IPs maintain connections with throttled Timed Sync frequencies, across \UniqueASNsthrottledTS unique ASs.



\subsubsection{Syntactic Anomalies \label{subsubsec:synanom}}
Beyond behavioral deviations, we applied \emph{SynAnom} (Sec.~\ref{subsubsec:synviol}) and identified three syntactic violations of Monero's protocol standard, indicating outdated or incorrectly implemented protocol standards. While not immediately critical in isolation, they are consistent with intentional customization and correlate with other anomalies identified above. 

\subparagraph{Support Flags Omission.} We identified nodes omitting the support flags field during Handshake exchanges. 
Support flags are enabled by default. These peers must have explicitly set this flag to 0 or falsely implemented a custom client. Beyond syntactic identification, this violation creates a unique command sequence pattern where support flags are explicitly requested through Support Flags request messages, followed by response messages, deviating from standard command sequence flows.
Empirically, we observed \totalUniqueIPssupportflag unique IPs omit the support flags, across \UniqueASNsSupportFlag unique ASs.

These missing flags resulted in \SFRpackets additional support flags request sequences, deviating standard protocol flow and creating increased network workload. 

\subparagraph{Deprecated \texttt{last\_seen} timestamps.} During PL exchanges in Handshake responses or Timed Sync responses, we observed peers still including the deprecated \texttt{last\_seen} timestamp field. 
As of our observation, the timestamp that these peers transmit appears to be placeholders fixed to a value, indicating an outdated custom implementation of a Monero node. Empirically, we observed that \totalUniqueIPslastseen unique IPs inside the ASLN continue to transmit the \texttt{last\_seen} timestamp. 


\subparagraph{} While these two anomalies represent syntactic violations, they have historically correct syntax and are still accepted in Monero's current connection handling. Nonetheless, these outdated definitions date back longer than the last network upgrade, which should have enforced every node to update their software, raising questions on the legitimacy of node software running on servers showing these violations.

\subparagraph{TCP Fragmentation.} Messages in P2P traffic can get bigger than the Maximum Transmission Unit (MTU) size. Those messages naturally get fragmented and reassembled and the size of the fragmentation can differ depending on the network path's MTU size. However, we identified IP addresses that consistently fragment packets in an unusual manner. For each reassembled message, the first packet only contains Monero's 8 byte signature. This persists even for messages that could be transmitted in a single packet. 
We observed \totalUniqueIPsSign unique IPs sending signature only packets, across \UniqueASNsSign unique ASs.


Such a behavior is unlikely to occur naturally, as legitimate MTU-based fragmentation would create more varied fragment sizes based on network conditions. The concentration within a small number of ASNs, including the ASLN, is consistent with coordinated deployment.

Individually, several of these anomalies admit benign explanations. Their joint co-occurrence within a shared AS, however, is not explained by hosting, VPN aggregation, or churn, and is most parsimoniously attributed to a single coordinated actor. We make this interpretation explicit in Sec.~\ref{sec:securityimpl}.

\section{From Anomalies to Adversary \label{sec:securityimpl}}
In this section, we first characterize the coordinated attack chain (Sec.~\ref{subsec:threatchain}), evaluate existing defenses (Sec.~\ref{subsec:defense}), and then propose mitigations (Sec.~\ref{subsec:mitigations}).
Throughout this section, we use the term "adversary" to refer to the entity or entities responsible for the anomalies characterized in Sec.~\ref{sec:monerosexamination}. We acknowledge in Sec.~\ref{sec:discussion} that anomalous behavior does not constitute proof of malicious intent. However, the scale, coordination, and network positioning of the observed anomalies support an adversarial interpretation. 

\subsection{A Coordinated Attack Chain \label{subsec:threatchain}}
\begin{figure}[t]
    \centering
    \includegraphics{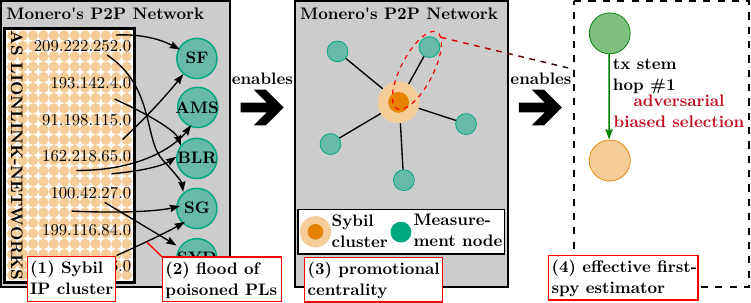}
    \caption{\textbf{ASLN Threat Chain.} AS LIONLINK operates seven densely populated subnets forming a Sybil cluster within Monero's P2P network. (1) The cluster's scale enables a Sybil attack, with the full saturation of /24 subnets suggesting that a small number of actual nodes operate behind proxy endpoints. (2) Peers across all seven subnets flood the network with poisoned PLs exclusively containing IP addresses from within these same subnets, as observed by each of our five monitoring nodes. (3) The combined effect of Sybil presence and PL poisoning translates into outsized promotional centrality, with adversarial peers consistently occupying a fraction of outgoing connections. (4) This central positioning enables an effective first-spy estimator: with high probability of being selected as an outbound relay during Dandelion++'s stem phase, the adversary can link transactions to their originating node.}
    \label{fig:threatchain}
  \end{figure}

\begin{figure}[t]
    \begin{subfigure}[t]{0.48\textwidth}
    \includegraphics[width=1\linewidth]{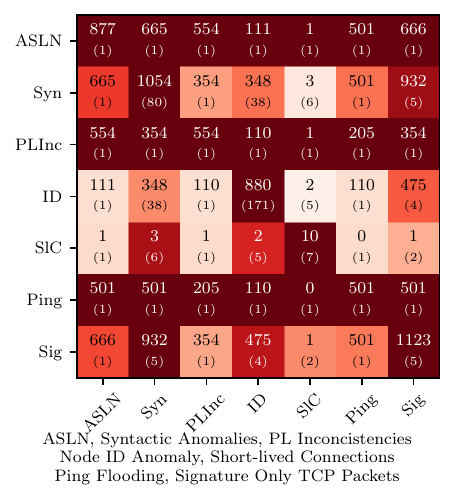}
    \caption{\textbf{Anomaly Overlap Matrix.} The matrix presents the overlap of IPs (ASs) showing anomalous behavior in the respective anomaly. Whereas many IPs are often affiliated with just a few ASs, some anomalies show higher AS overlap than IP overlap, suggesting coordinated deployment of peers with distinct functionalities.}
    \label{fig:anomalyOverlap}
  \end{subfigure}
  \hfill
  \begin{subfigure}[t]{0.48\textwidth}
    \includegraphics[width=1\linewidth]{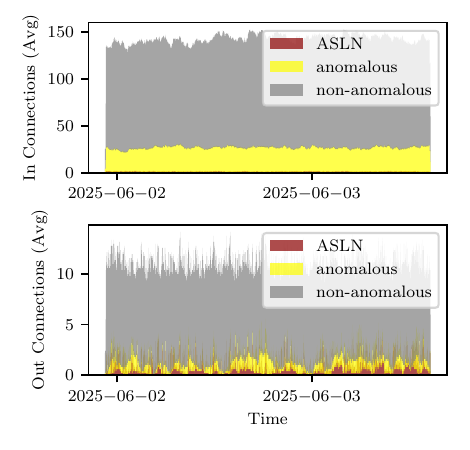}
    \caption{\textbf{Proportion of Active Connection Saturation.} The graph visualizes the proportion of active connections. Outgoing and incoming connections show on average around \outgoingConnSatwithout\% and \incomingConnSatwithout\% are occupied by non-standard peers and \outgoingConnSatAdvwithout\% and \incomingConnSatAdvwithout\% are occupied by adversarial peers.}
    \label{fig:connSaturation}
  \end{subfigure}
  \caption{\textbf{Adversarial Reach.} The overlap matrix (left) characterizes which anomalies share IP and AS affiliation. The connection saturation graph (right) quantifies how this translates into direct presence within an individual node's active connection pool.}
\end{figure}

The adversary's operational footprint likely extends beyond ASLN. As shown in Figure~\ref{fig:anomalyOverlap}, the anomaly overlap reveals that many identified anomalies extend beyond ASLN into additional ASs. This suggests the adversary operates nodes with distinct IP addresses outside their primary infrastructure, either as auxiliary capacity or to reduce detectability. 
We focus our threat analysis on the ASLN cluster (1,582 nodes), where anomalies are most concentrated and corroborated, but the true operational footprint is likely broader. We break down the threat chain shown in Figure~\ref{fig:threatchain} by evaluating the identified anomalies and explaining the effect each enables. 
\subparagraph{(1) Sybil attack.} At the center of our analysis lies a set of seven highly saturated subnets. The concentration of 254 peers within a single /24 subnet is strongly indicative of a Sybil attack, a conclusion further supported by several corroborating anomalies. First, these peers belong to large peer ID clusters characterized by shared and frequently rotated identifiers. Second, they exhibit a distinctive communication pattern consisting of signature-only packets. This uniform behavior suggests these IPs function as proxy or VPN endpoints load-balancing traffic for a limited number of actual Monero nodes.

\subparagraph{(2) Peer List Poisoning.} The semantic PL inconsistencies imply poisoning of the discovery mechanism to increase visibility and gain strategic positioning. Although nodes will prefer to choose outgoing connections to distinct /16 subnets, high amounts of nodes in the local white and gray lists will naturally increase the chance of picking a node in the ASLN.
This visible PL manipulation occurs exclusively for the ASLN but appears to be very effective. The Sybil attack combined with the PL poisoning results in widespread exposure across the network. We examined the inferred direct exposure of each node by analyzing all collected PLs along their promoted peers. Our measurements reveal that the majority of PLs we received contain ASLN peers, and a peer's PL is on average occupied by \avgPeerPoisoning\% ASLN peers.

\subparagraph{(3) Central Network Positioning.} The promotion topology analysis in Sec.~\ref{subsec:promtop} establishes that ASLN achieves outsized promotional prominence in our captured PLs. To quantify how this translates into direct exposure on a node-level, we measured how many ASLN and anomalous peers occupy our measurements' nodes active connection pools over the observation period, averaged across all five nodes.
As shown in Figure~\ref{fig:connSaturation}, ASLN peers occupy \incomingConnSatAdvwithout\% of incoming and \outgoingConnSatAdvwithout\% of outgoing connections on average. Additionally, \incomingConnSatwithout\% and \outgoingConnSatwithout\% of incoming and outgoing connections are occupied by anomalous peers. This degree of saturation is not coincidental but a direct operational consequence of the Sybil attack and PL poisoning described in steps (1) and (2). By flooding the network with nodes and ensuring those nodes dominate peer lists, the adversary translates promotional influence into persistent connection-level presence. This central positioning is the precondition for the privacy attacks described next.

\subparagraph{(4) Transaction Privacy Degradation.} The presence and network positioning of ASLN peers creates the preconditions necessary to compromise Dandelion++. The main capability to undermine privacy guarantees is based on the central position its nodes achieve through effective PL poisoning and Sybil attacks. Dandelion++ applies an initial stem phase where the transaction is only forwarded to one of the node's chosen outbound relays. 
With the ASLN occupying on average \outgoingConnSatAdvwithout\% of each node's outgoing connection pool, the adversary consistently has a chance of being picked as an outbound relay, thereby being positioned to learn about new transactions and apply a first-spy estimator (a deanonymization technique linking the first node to relay a transaction as its likely originator). Assuming the ASLN successfully learns the network's topology, it can further limit possible originators beyond relying solely on the first-spy-estimator. According to Sharma et al.~\cite{sharma2022anonymity}, an adversary that controls around 20\% of the network can intercept around 70\% of transactions and effectively reduce the anonymity set to 32 possible originators per transaction. However, we observed a node's connection pool to be occupied by \outgoingConnSatwithout\% of anomalous peers. This scenario potentially constitutes a more critical threat model than conventional network distribution approaches. Widespread connection saturation may enable direct node identification or, as approached by Sharma et al., allow adversaries to systematically eliminate candidate nodes, constraining the anonymity set to a limited subset of potential originators.

Finally, in certain cases, the adversary might confidently identify a transaction's originator. If the adversary knows whether a node is reachable (i.e., whether it allows inbound connections) and receives a stem transaction from such an unreachable node, it can confidently identify this as the originator since the node cannot have relayed the transaction from elsewhere.

\subparagraph{Eclipse Attacks.} Another final attack mapping follows a combined approach of Handshake and Ping flooding as reported in two distinct vulnerability reports~\footnote{https://hackerone.com/reports/3185083, https://hackerone.com/reports/2858802}. The observed patterns also align closely with the eclipse attack methodology described by Shi et al.~\cite{shi2025eclipse}. Many distinct peers align around each measurement node, able to quickly fill a victim's gray and white lists once needed. The next step requires a connection reset attack. While the double-spend attack vector has been closed, we observed many attempts to perform Ping flooding attacks, which have a similar outcome but have also been disclosed and fixed. We did not observe a successful eclipse attack against our measurement nodes. Our observations are most consistent with a broad monitoring and Sybil deployment rather than a targeted eclipse campaign. A targeted eclipse attacker would concentrate resources on selected victims and would largely avoid simultaneous exposure at five unrelated vantage points. Our data therefore speaks to eclipse capability rather than to an attempt in progress.

\subsection{Defense Mechanisms \label{subsec:defense}}
As a potential countermeasure, Monero contributors advertise the usage of a community crafted ban list. We evaluate the effectiveness of this defense against the identified threats.

\subparagraph{Ban List Effectiveness}
To compare the ban list with our identified anomalous nodes, we merge all anomalous peers identified across measurement nodes, arriving at 1,877 anomalous IP addresses. Of these, 734 are not in the community proposed ban list. With an active ban list and an increased pool of outgoing connections, we observed no Ping flooding attacks and outgoing connections occupied by anomalous peers dropped to an average of \outgoingConnSatwith\%. The ASLN is fully blocked on the ban list. 

\subparagraph{Ban List Limitations.} Although a ban list seems to increase privacy from an individual node's perspective, we observed ban-listed peers being promoted in the majority of PLs. This suggests that many nodes in the network have not enabled the recommended ban list. As such nodes will be neighbors of nodes having a ban list enabled, the adversary will most likely be able to map these neighboring nodes despite their active ban list. Extending this, even if a majority is reached that have enabled an effective ban list, updates must be maintained and monitored on an ongoing basis. The static nature of a ban list cannot adapt to a rapidly evolving adversary. 

Furthermore, the results showed that even with an active ban list many distinct anomalies persisted. We continued to observe several distinct anomalous behaviors for each measurement node with an active ban list throughout the observation period. Extending ban list coverage to entire ASs would reduce exposure further, but an adversary with sufficient resources to deploy a diverse, multi-AS infrastructure can circumvent AS-level filtering by design. The fundamental limitation is architectural, as a static, externally-maintained list reacts to observed behavior rather than preventing it.  

\subsection{Mitigation Strategies\label{subsec:mitigations}}
Protocol changes such as the removal of the \texttt{last\_seen} timestamp or support flags date back to 2018, but the network does not yet punish false usage of either. The already existing offense tracker could be updated to account for syntactic incorrectness and act accordingly. 

The effectiveness of PL poisoning could be reduced through periodic pruning of a node's local white and gray lists, removing duplicated subnets or even ASs. In addition, a set of anchor nodes, potentially the seed nodes, could act as a regular peer, providing a healthy PL to help counteract the self-reinforcing nature of PL contamination.

However, these mitigations address symptoms rather than root causes, and each defense carries its own tradeoffs. Validating PL entries and pruning duplicated subnets raises the cost of PL poisoning, but favoring older or more stable peers risks introducing centralization and ossification, and constraining peer selection too far erodes the randomness of outgoing connections on which eclipse resistance depends. Reputation systems for peer behavior can themselves be accumulated and gamed by Sybil identities, shifting the attack surface rather than eliminating it. Any effective mechanism to detect and penalize coordinated Sybil networks must therefore be designed with these manipulation risks in mind. The sophistication of the observed adversary suggests that incremental improvements to existing defenses may prove insufficient without fundamental architectural changes to the peer discovery and connection management mechanisms, yet such changes must balance Sybil resistance against decentralization and openness. The community ban list remains a useful short-term measure but provides no durable security guarantee against the adversary characterized in this paper.

Complementing these protocol-level changes, our released pipeline lets operators act today: it takes a PCAP captured by their own node and outputs a per-anomaly breakdown, reporting each flagged IP alongside the specific anomaly classes it triggered. Operators can inspect \emph{why} a peer was flagged rather than trusting an opaque list, and derive a local ban list from the output. As it is regenerated from current traffic, the breakdown adapts to redeployed infrastructure, addressing the static ban list's central limitation (Sec.~\ref{subsec:defense}). Flags remain indicators rather than proof of malice, so operators tune sensitivity against their review capacity.


\section{Discussion \label{sec:discussion}}
Our approach provides us with a verified behavioral baseline from five geographically distributed locations. Running the default and unmodified Monero software allows us to distinguish standard from anomalous behavior, and to infer non-standard implementations where the observed behavior cannot be produced by an unmodified Monero client. This consistent baseline behavior, combined with Monero's network architecture presuming homogeneous node software deployment, ultimately allows us to distinguish between standard and non-standard behavior.

\subparagraph{Thresholds, Anomalies, and Malicious Intent.} We identified many IPs showing anomalous signs, associated with either few or many distinct ASs. While syntactic anomalies are clear indicators, behavioral inconsistencies are more ambiguous. Node ID anomalies, throttled Timed Syncs, or PL similarities might indicate anomalous peers without malicious intent. Several benign explanations exist for individual anomalies in isolation. Shared VPN endpoints may cause multiple IPs to announce the same peer ID. Network congestion may account for throttled Timed Sync frequencies. Peering agreements between hosting providers could produce correlated peer lists over time. Short-lived connection bursts and elevated Ping rates are likewise characteristic of active measurement infrastructure. However, these explanations do not account for the joint co-occurrence of anomalies within a single AS, nor for the scale and uniformity of the observed behavior.
While our node's data serves as a verified baseline for identifying anomalies, direct classification as malicious requires verified labels. However, consistent deviation from network baseline remains significant for security contexts, regardless of ultimate intent.
A sensitive threshold can produce many false positives requiring manual review but reducing missed threats. An insensitive threshold can prioritize precision at the cost of potentially overlooking subtle anomalies. In our examination, we focused on clear outliers in our dataset rather than defining specific thresholds, as optimal values depend on operational capacity and use case requirements.

\subparagraph{Community Awareness.} Monero's community has been increasingly aware of potential surveillance and malicious activities within the network with an ongoing discussion and monitoring\footnote{https://moneronet.info} that partially overlap with our findings. Most prominently, repeated ID switching is used to fingerprint spy nodes, and affected IPs are added to a continuously updated ban list. Our anomaly detection approach complements these community efforts by providing a comprehensive framework with empirical evidence for identifying non-standard nodes. The Monero Research Lab is aware of our analysis and already is testing our examination pipeline on newly captured data, which supports the practical applicability of our approach despite the absence of formal ground truth data.

\section{Conclusion \& Future Work}
In this work, we have proposed methods to identify anomalous nodes in Monero's P2P network. We have found that approximately \percentageAnomNet\% of the observed peers in Monero's P2P network exhibit anomalous behavior. At the time of measurement, this indicated that a well-resourced actor was active in the network, performing what appears to be PL poisoning, topology mapping, and connection saturation. Despite community mitigation efforts, including when adopting ban lists, anomalous peers still occupied \outgoingConnSatwith\% of outgoing connections, potentially undermining Monero's privacy guarantees and creating a critical gap between theoretical protections and practical vulnerabilities.

First, we envision further study of the promotion topology as we have identified interesting patterns that might yield further insights. Second, another direction concerns the impact on Dandelion++ considering adversarial structures which go beyond random distributions in the network.




\bibliography{main}

\appendix

\end{document}